\PassOptionsToPackage{pdfpagelabels=false}{hyperref}
\documentclass[fleqn,usenatbib]{mnras}

\usepackage[T1]{fontenc}
\usepackage{ae,aecompl}
\usepackage{float}

\usepackage{graphicx}	
\usepackage{amsmath}	
\usepackage{amssymb}	
\usepackage{bm}         
\usepackage[usenames]{color}     
\usepackage{xspace}

\usepackage{epstopdf}

\usepackage{xcolor}
\definecolor{green}{HTML}{33CC33}
\definecolor{red}{HTML}{FF3300}
\definecolor{blue}{HTML}{3333FF}

\usepackage{threeparttable}
\usepackage{booktabs}

\usepackage[english]{babel}

\renewcommand{\eqref}[1]{Equation~\ref{#1}}
\newcommand{\fref}[1]{Figure~\ref{#1}}
\newcommand{\tref}[1]{Table~\ref{#1}}
\newcommand{\sref}[1]{Section~\ref{#1}}

\newcommand{\ie}{i.e.\@\xspace} 
\newcommand{\eg}{e.g.\@\xspace} 

\newcommand{\numax}{\ensuremath{\nu_{\rm max}}\xspace}

\newcommand{\kp}{\textit{Kepler}\xspace}
\newcommand{\co}{CoRoT\xspace}

\newcommand{\teff}{\ensuremath{T_{\rm eff}}\xspace}
\newcommand{\logg}{\ensuremath{\log g}\xspace}
\newcommand{\feh}{\ensuremath{\rm [Fe/H]}\xspace}
\newcommand{\meh}{\ensuremath{\rm [M/H]}\xspace}
\newcommand{\cp}{\ensuremath{c_{P\rm -bol}}\xspace}

\numberwithin{equation}{section}

\makeatletter
\def\maketag@@@#1{\hbox{\m@th\normalfont\normalsize#1}}
\makeatother

\makeatletter
\DeclareRobustCommand*\textsubscript[1]{%
  \@textsubscript{\selectfont#1}}
\def\@textsubscript#1{%
  {\m@th\ensuremath{_{\mbox{\fontsize\sf@size\z@#1}}}}}
\makeatother

\makeatletter
\newcommand*\mysize{%
  \@setfontsize\mysize{5.7}{8.0}%
}
\makeatother

\makeatletter
\newcommand*\tabsize{%
  \@setfontsize\tabsize{7.}{8.0}%
}
\makeatother

\usepackage{blindtext}
\usepackage{lipsum} 

\defcitealias{B11}{B11} 
\defcitealias{M09}{M09}

\usepackage[labelsep=endash,font=footnotesize]{subfig}

\title[Bolometric correction for solar-like oscillations]{Bolometric corrections of stellar oscillation amplitudes as observed by the \kp, \co, and TESS missions}

\author[Lund, M. N.]{Mikkel~N.~Lund$^{1,2}$\thanks{E-mail: \href{mailto:mikkelnl@phys.au.dk}{mikkelnl@phys.au.dk}}
\vspace*{0.5em} \\ 
$^1$Stellar Astrophysics Centre, Department of Physics and Astronomy, Aarhus University, Ny Munkegade 120, DK-8000 Aarhus C, Denmark\\
$^2$School of Physics and Astronomy, University of Birmingham, Edgbaston, Birmingham, B15 2TT, UK\\
}

\begin{document}

\vspace{-10cm}
\date{Accepted 2019 July 2; Received 2019 July 2; in original form 2019 March 14}

\pagerange{\pageref{firstpage}--\pageref{lastpage}} \pubyear{2019}

\maketitle

\label{firstpage}
\vspace{-3cm}
\begin{abstract}
{A better understanding of the amplitudes of stellar oscillation modes and surface granulation is essential for improving theories of mode physics and the properties of the outer convection zone of solar-like stars. A proper prediction of these amplitudes is also essential for appraising the detectability of solar-like oscillations for asteroseismic analysis.
Comparisons with models, or between different photometric missions, are enabled by applying a bolometric correction, which converts mission-specific amplitudes to their corresponding bolometric (full light) values.} 
{We derive the bolometric correction factor for amplitudes of radial oscillation modes and surface granulation as observed by the \kp, \co, and TESS missions. The calculations are done assuming a stellar spectrum given by a black-body as well as by synthetic spectral flux densities from 1D model atmospheres. We derive a power-law and polynomial relations for the bolometric correction as a function of temperature from the black-body approximation and evaluate the deviations from adopting a more realistic spectrum.} 
{Across the full temperature range from $4000-7500$ K, the amplitudes from TESS are in the black-body approximation predicted to be a factor ${\sim}0.83-0.84$ times those observed by \kp. We find that using more realistic flux spectra over the black-body approximation can change the bolometric correction by as much as ${\sim}30\%$ at the lowest temperatures, but with a change typically within ${\sim}5-10 \%$ around a \teff of $5500-6000$ K. We find that after \teff, the bolometric correction most strongly depends on \meh, which could have an impact on reported metallicity dependencies of amplitudes reported in the literature.} {}
\end{abstract}



\section{Introduction}
The study of the physical processes underlying the interaction between pulsations and convection \citep[][]{2015LRSP...12....8H} requires observed values of oscillation mode amplitudes. Space-based missions such as \kp have lead to a large number of such amplitude measurements, including derivations of amplitude relations as a function of fundamental stellar parameters, such as luminosity and effective temperature \citep[see, \eg,][for such relations from \kp observations]{2011ApJ...743..143H,2012A&A...537A..30M,2013MNRAS.430.2313C}.

However, the general use of the measured oscillation mode amplitudes requires a conversions between what is observed with a given spacecraft to the corresponding bolometric (full light) value. This conversion is given by the bolometric correction \cp for radial ($l=0$) modes of oscillation, which can also be applied to estimate the bolometric intensity fluctuation from granulation \citep[see][]{Kallinger2014}. Such a bolometric correction was calculated by \citet{B11} (hereafter \citetalias{B11}) and \citet{M09} (hereafter \citetalias{M09}) for the observing band-pass of the \kp \citep{VCleve} and \co \citep[][]{2009A&A...506..411A} missions.

The bolometric correction is thus also essential for applying amplitude relations derived for one mission to another. Such a use is especially important for predicting the detectability of solar-like oscillations and thereby guide the target selection strategy for a mission such as TESS \citep[][]{2011ApJ...732...54C,2016ApJ...830..138C,2019arXiv190110148S}. 

\begin{figure}
\includegraphics[width=\columnwidth]{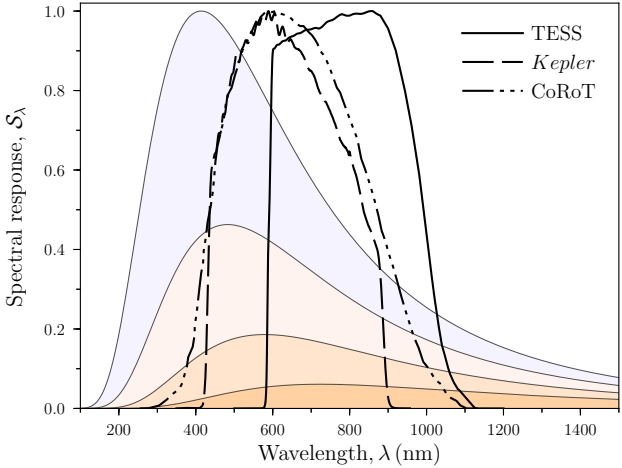}
\caption{Spectral response functions $\mathcal{S}_{\lambda}$ for the \kp \citep{VCleve}, \co \citep{2009A&A...506..411A}, and TESS \citep{2014SPIE.9143E..20R} missions as a function of wavelength $\lambda$ (normalised to a maximum of $1$). The shaded regions show black-body spectra with temperatures of $7000, 6000, 5000$, and $4000$ K (normalised to a maximum of $1$ for the hottest/bluest curve).}
\label{fig:Transfer_functions}
\end{figure}

In this paper we calculate the bolometric correction \cp for the photometric missions \kp, \co, and TESS \citep{2014SPIE.9143E..20R}.
In \sref{sec:bol} we outline the procedure for the calculation. The calculations of \cp are presented in \sref{sec:bol1}, both from using a Planck spectrum in \sref{sec:bol2} and from adopting synthetic model flux spectra in \sref{sec:bol3}. In \sref{sec:dis} we discuss the potential impact from changes to the bolometric correction, and conclude in \sref{sec:con}.
\begin{figure*}
    \centering
    \subfloat{\includegraphics[width=0.47\textwidth]{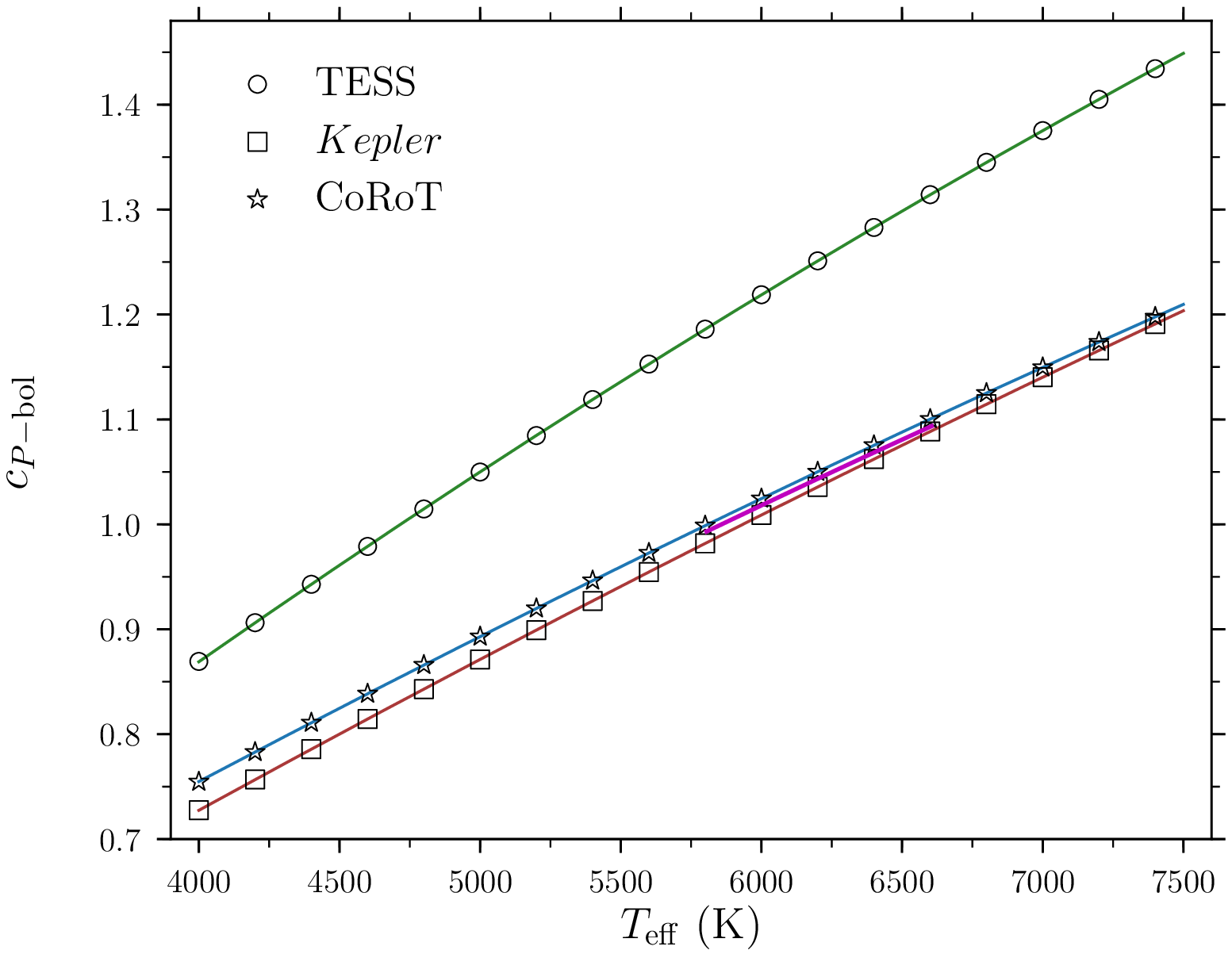}}
    \hfill
    \subfloat{\includegraphics[width=0.47\textwidth]{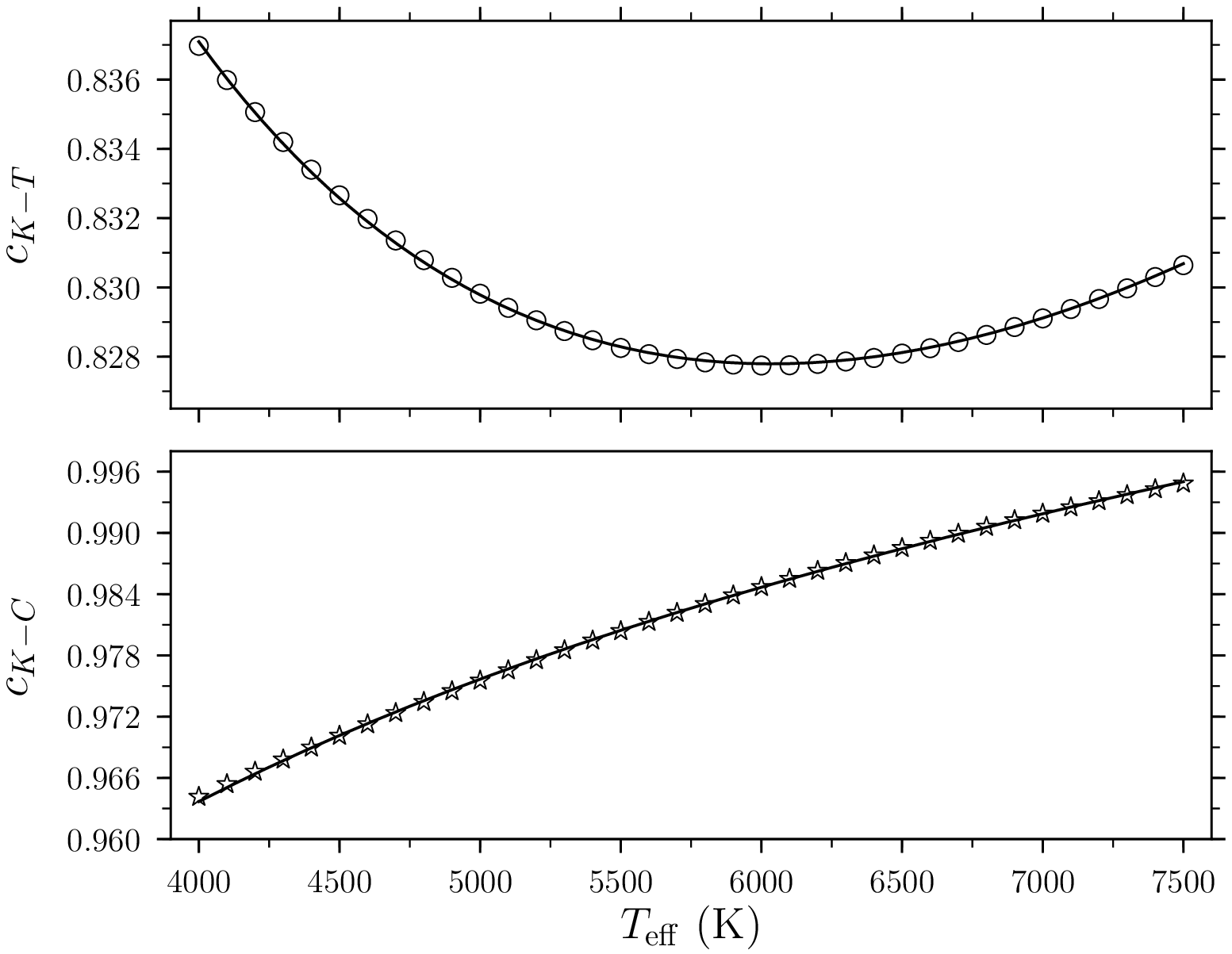}}
\caption{Left: Calculated values for the bolometric correction \cp (\eqref{eq:cbol}), from adopting a Planck spectrum (\eqref{eq:black}), as a function of \teff. Values are given for both the \kp, TESS, and \co spectral response functions (see legend). The lines show the $T_2$ (TESS), $K_2$ (\kp), and $C_2$ (\co) polynomial fits (\eqref{eq:rel2}), with coefficients provided in \tref{tab:rel_coeff}. The magenta line from \teff $5800-6750$ K gives the relation obtained by \citetalias{M09} from fitting to calculated $c_{C\rm -bol}$ values in this \teff range, but with values obtained from ATLAS9 model atmosphere fluxes covering a small range in \logg and \feh around the Solar value.
Right: Ratio between the bolometric corrections from the left panel for \kp and TESS ($c_{K-T}$; \eqref{eq:conv}), and \kp and \co ($c_{K-C}$; \eqref{eq:conv}). The line gives the rational function between the $K_2$ and $T_2$ models for $c_{K-T}$, and $C_2$ and $T_2$ models for $c_{K-C}$, given by the ratio of these respective relations (\tref{tab:rel_coeff}).}
	\label{fig:relation}
\end{figure*}    


\section{Bolometric correction \cp}\label{sec:bol}
The bolometric correction \cp is defined such that it gives the conversion factor between bolometric amplitudes and those observed through a given band-pass `$P$': 
\begin{equation}\label{eq:Bbol}
A_{\rm bol} = \cp\,  A_{P}\, .
\end{equation}
Throughout, when referring specifically to either the \kp, \co, or TESS band-pass we shall replace `$P$' by `$K$', `$C$', or `$T$' respectively. Hence, the bolometric correction for TESS reads `$c_{T\rm -bol}$'. 

Following the derivation by \citetalias{B11} and \citetalias{M09} \citep[see also][]{1990A&A...227..563B} for the \kp and \co missions, \cp may be approximated as\footnote{This relates to the response function  $R_g$ of \citetalias{M09} as $c_{P\rm -bol} = 4/R_g$.}:
\begin{equation}\label{eq:cbol}
c_{P\rm -bol} = \frac{4 \int \tau_P(\lambda) F(\lambda,\cdot)\, \mathrm{d}\lambda }{\teff \int \tau_P(\lambda) \frac{\partial F(\lambda,\cdot)}{\partial\teff}  \,\mathrm{d}\lambda  }\, .
\end{equation}
Here, $F(\lambda,\cdot)$ is the representation of the stellar spectral flux density; $F(\lambda,\cdot)$ can be given by the true (or simulated) spectral flux density, in which case it will depend on several parameters, \eg, $F(\lambda,\cdot) = F(\lambda,\teff, \logg, \feh,\cdot)$, or be represented by a black-body (Planck) function, \ie, $F(\lambda,\cdot) = B(\lambda,\teff)$ (as done in \citetalias{B11}). In this work the bolometric corrections from both of these options are calculated. 
For reference, $B(\lambda,\teff)$ in flux density units is given as:
\begin{equation}\label{eq:black}
B(\lambda,\teff) =  \frac{2\pi h c^2}{\lambda^5\left(e^{\frac{hc}{\lambda\, k_{\rm B} \teff }}-1\right)}\,\,\, \left[\rm erg\, cm^{-3}\, s^{-1} \right],
\end{equation}
where $\lambda$ gives the wavelength, $h$ is Planck's constant, $c$ is the speed of light, and $k_{\rm B}$ is Boltzmann's constant. We note the added value of $\pi$ to convert from spectral radiance to spectral flux density.
The temperature derivative of the black-body function is given as:
\begin{equation}\displaystyle
\frac{\partial B(\lambda,\teff)}{\partial\teff} = \frac{2\pi h^2c^3 e^{\frac{hc}{\lambda\, k_{\rm B} \teff }}}{\lambda^6\, k_{\rm B}\, \teff^2 \left(e^{\frac{hc}{\lambda\, k_{\rm B} \teff }}-1\right)^2}\,\,\, \left[\rm erg\, cm^{-3}\, s^{-1}\,K^{-1}\right].
\end{equation}
In \eqref{eq:cbol}, $\tau_P(\lambda)$ is the instrumental transfer function for the band-pass `$P$' given as:
\begin{equation}
\tau_P(\lambda) = \mathcal{S}_{\lambda}/E_{\lambda}\,\,\, \left[\rm erg^{-1}\right] ,
\end{equation}
where $E_{\lambda}=hc/\lambda$ is the photon energy at wavelength $\lambda$, and $\mathcal{S}_{\lambda}$ is the spectral response function (SRF) defining the band-pass throughput.

In \fref{fig:Transfer_functions} we show the SRFs for \kp, \co, and TESS as a function of $\lambda$. 
The TESS SRF can be seen as the product of a long-pass filter transmission at short wavelengths and the quantum efficiency (QE) of the detector at long wavelengths. The TESS band-pass is roughly centred on the Johnson-Cousins $I_C$ filter at ${\sim}800$ nm \citep{2014SPIE.9143E..20R,2015ApJ...809...77S}, and it is redder than that of \kp, the latter being is closer to the $R_C$ filter \citep[][]{2010ApJ...713L..79K}. The \co band-pass is very similar to that of \kp, but slightly wider, especially towards long wavelengths.


\section{\cp relations}\label{sec:bol1}

\subsection{\cp from a black body approximation}\label{sec:bol2}

\fref{fig:relation} (left panel) shows calculated values for \cp (\eqref{eq:cbol}) for the three missions as a function of \teff in the range from $4000 - 7500$ K. The numerical integration of the components of \eqref{eq:cbol} were computed using a tanh-sinh quadrature scheme as implemented in the \texttt{Python} module \texttt{mpmath} \citep{mpmath}.
Like \citetalias{B11}, we fitted two types of parameterizations to the calculated values, viz., a power-law:
\begin{equation}\label{eq:rel1}
c_{P\rm -bol}(\teff) \approx \left(\frac{\teff}{T_o}\right)^{\alpha}\, ,
\end{equation}
and a polynomial:
\begin{equation}\label{eq:rel2}
c_{P\rm -bol}(\teff) \approx \sum_{i=0}^4 a_i(\teff-T_o)^{i}\, ,
\end{equation}
which we fitted including 2 orders.
The coefficients from the different fits are provided in \tref{tab:rel_coeff} along with the root-mean-square error ($\sigma_{\rm rms}$) to indicate the `goodness-of-fit'.
For TESS the power-law model ($T_1$; \eqref{eq:rel1}) is found to perform worse than the corresponding for \kp and \co, but still reproduces the $c_{T\rm -bol}$ values with residuals within $\pm3\times 10^{-3}$. The polynomial parameterizations have residuals within about $\pm1\times 10^{-4}$. 

To estimate the conversion factor between amplitudes observed in \kp, TESS, and \co one can use the rational function $\mathcal{R}(\teff)$, given by the ratio between the relations describing $c_{P\rm -bol}(\teff)$ for the different missions. For \kp versus TESS, $\mathcal{R}(\teff)$ thus provides an approximate relation for:
\begin{equation}\label{eq:conv}
c_{K-T}(\teff)=  \frac{c_{K\rm -bol}(\teff)}{c_{ T\rm-bol}(\teff)}\, .
\end{equation}
\fref{fig:relation} (right panel) shows the calculated values of $c_{K -T}(\teff)$ and $c_{K -C}(\teff)$ together with the rational functions $\mathcal{R}(\teff)=K_2 / T_2$ and $\mathcal{R}(\teff)=K_2 / C_2$. As seen, for TESS the conversion factor ranges between ${\sim}0.83$ and ${\sim}0.84$ in the temperature range considered, thus on average the amplitudes observed by TESS will be ${\sim}83-84\%$ that observed in \kp. For \co the amplitudes are very similar to those of \kp, as one might expect from the very similar spectral response functions (\fref{fig:Transfer_functions}), ranging between ${\sim}0.966$ and ${\sim}0.996$.


\subsection{\cp from synthetic spectra}\label{sec:bol3}
\begin{figure}
\includegraphics[width=\columnwidth]{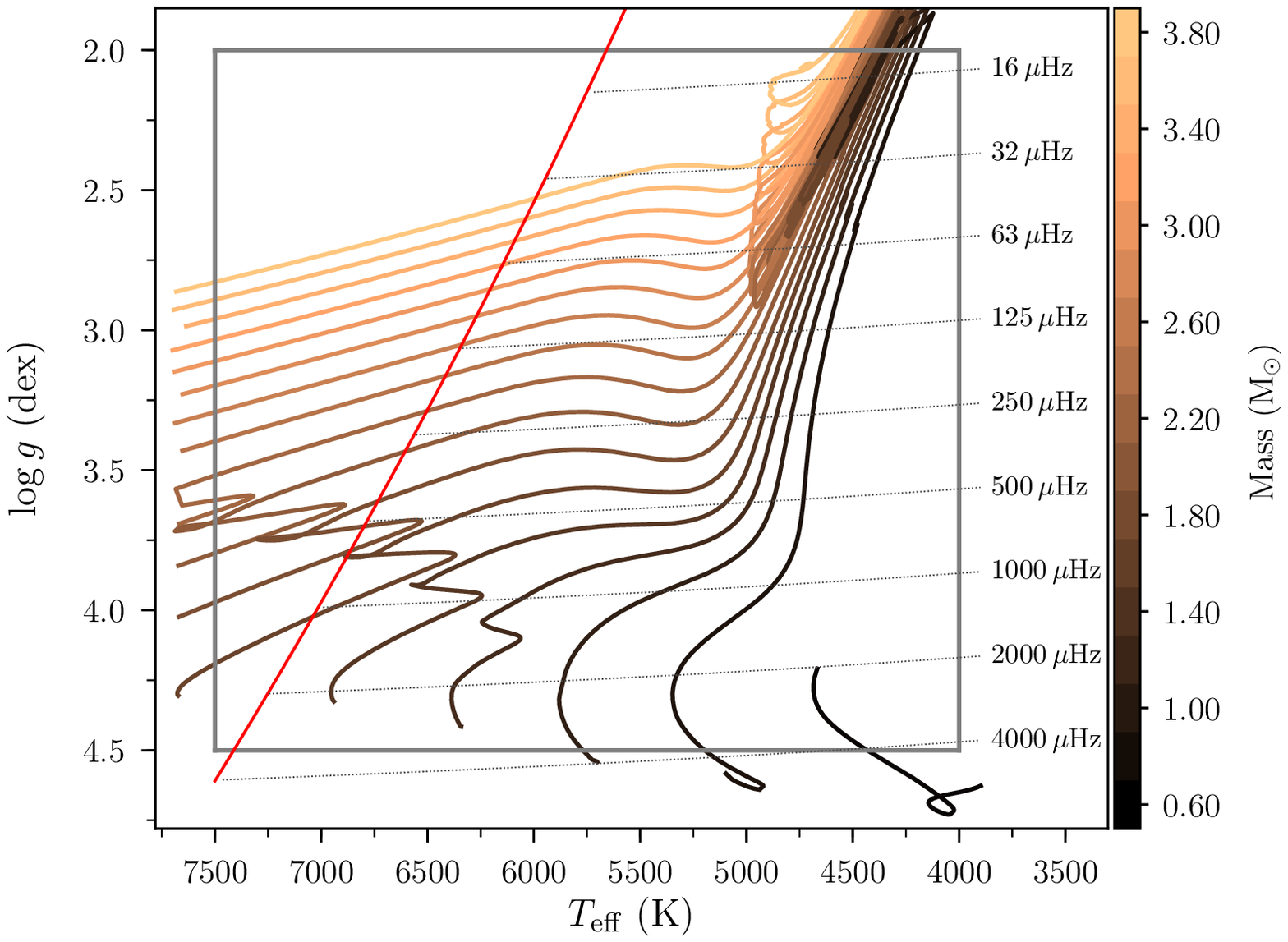}
\caption[]{Kiel-diagram illustrating by full grey lines the boundaries of the bolometric correction calculations. Stellar evolution tracks are obtained from MIST\footnotemark \citep{2016ApJ...823..102C}, adopting $\feh=0$ (using the \citet{2009ARA&A..47..481A} solar reference) and a range of masses as indicated by the colour-bar. The red line gives the red-edge of the classical instability strip from \citet{2000ASPC..210..215P}, to the left of which no solar-like oscillations are expected. Indicated are also lines of constant \numax, the frequency of maximum oscillation mode amplitude.}
\label{fig:evol}
\end{figure}
\footnotetext{\url{http://waps.cfa.harvard.edu/MIST/}}
To investigate the impact from using the simplistic representations for the stellar flux density given by the Planck spectrum, we computed \cp values using a grid of ATLAS9\footnote{\url{http://wwwuser.oats.inaf.it/castelli/grids.html}} model fluxes. All spectra are computed with updated opacity distribution functions \citep[ODFNEW;][]{2003IAUS..210P.A20C}, without overshoot, assuming a mixing length of $\ell/H_p=1.25$, a microturbulence of $2\, \rm km/s$, and adopting the \citet{Grevesse:1998cy} reference for the solar metal content.  

For these calculations of \cp we used spectra with temperatures from $4000-7500\, \rm K$ in steps of $250\,\rm K$. For each temperature we further adopted surface gravity ($\log\, g$) values from $2.0$-$4.5$ in steps of $0.5$ dex, and metallicities of $\rm [M/H]=-1, 0, +0.2$ and $+0.5$ dex. For all combinations of the above we further adopt two different values for the $\alpha$-enhancement, namely, $\rm [\alpha/M]=0$ and $\rm [\alpha/M]=+0.4$. \fref{fig:evol} illustrates in a Kiel-diagram the region covered by the calculations. As seen, some combinations covered by the calculation will not be realistic for a real star or correspond to stars showing solar-like oscillations. The calculations do, however, cover most regions populated by detected solar-like oscillators from the \kp and \co missions \citep[see, \eg,][]{2011ApJ...743..143H,2013ARA&A..51..353C}.
\begin{figure*}
\includegraphics[width=\textwidth]{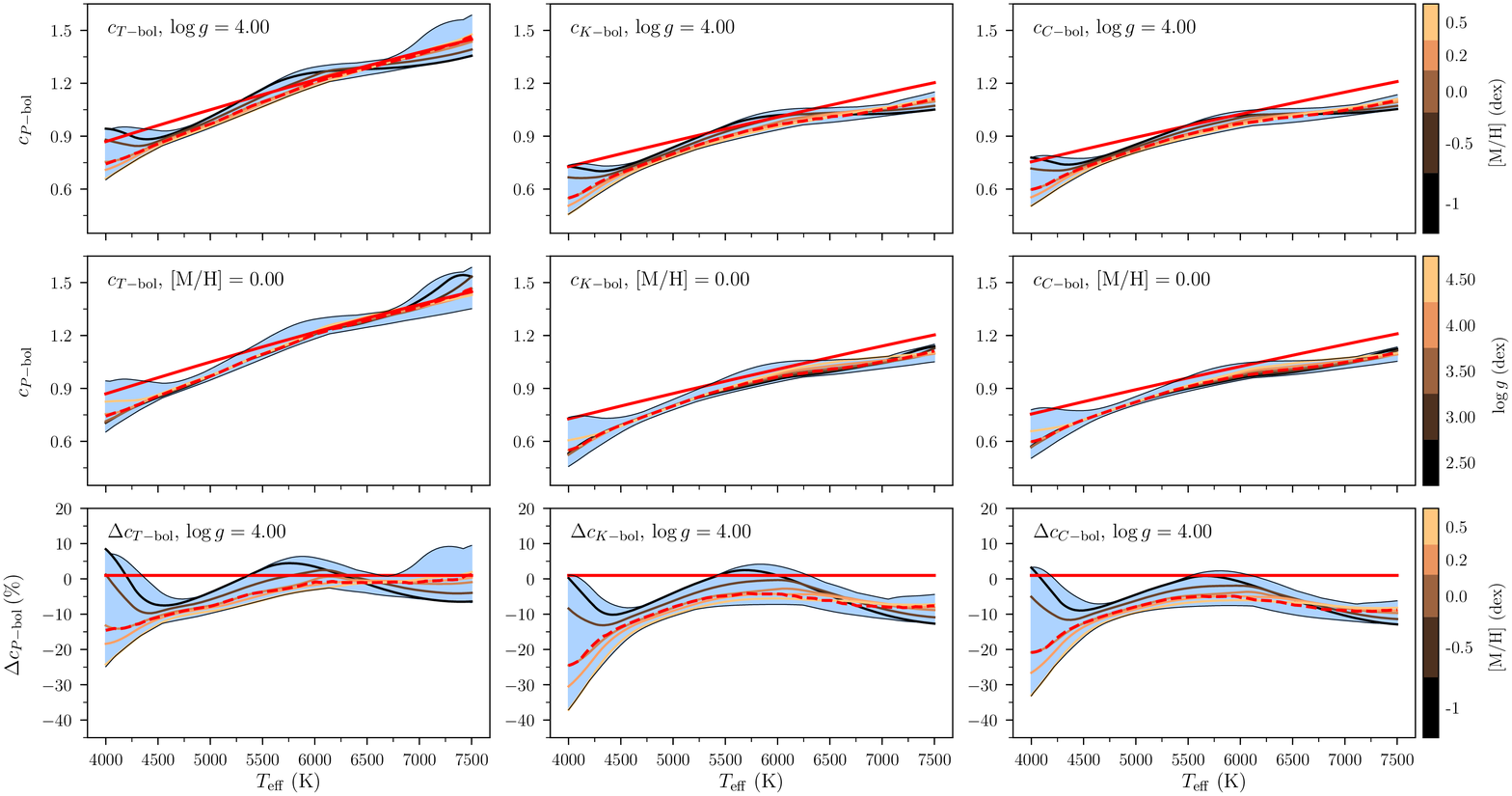}
\caption{Calculated values for \cp as a function of \teff. Left column panels show values for TESS, middle column panels for \kp, and right column panels for \co. The lines coloured according to the colour-bar give \cp for a specific \logg and a range of \meh values (top row), or a specific \meh and a range of \logg values (middle row). In all panels the full red line gives the \cp value from assuming a Planck spectrum; the blue shaded area shows the regions covered by all combinations of \logg and \meh, and the dashed red line gives the median \cp value for all these combination. The bottom row give the percentage difference between the Planck and synthetic spectrum \cp values, with lines adopted from the top row.}
\label{fig:Bol_from_spec_ex}
\end{figure*}
\fref{fig:Bol_from_spec_ex} shows examples of the calculated \cp values for the TESS missions for a given \logg in combination with all \meh (top row), and for a given \meh in combination with all \logg values (middle row). From these examples it is evident that (1) differences in \cp are mainly a function of metallicity, while \logg has little impact; (2) the values computed using more realistic flux densities are typically lower than those obtained using the simple Planck spectrum; (3) the best agreement between the two \cp estimates is typically found around \teff values of $5500-6500$ K. For TESS the absolute deviations from the Planck estimates (\fref{fig:Bol_from_spec_ex}; bottom row) reaches at the lowest temperatures and highest metallicities up to ${\sim}30\%$. 
\fref{fig:Bol_from_comp} shows the resulting rational functions for the different missions (top row) and percentage deviation from the Planck spectrum rational functions (bottom row). Because the spectrum-derived \cp for the different missions overall follow the same behaviour with \teff, \logg, and \meh, the ratio between two such values shows a smaller difference to the Planck-derived ratio than does the individual \cp values. The absolute difference to the Planck rational function reaches up to ${\sim}15\%$.
\begin{figure*}
\includegraphics[width=\textwidth]{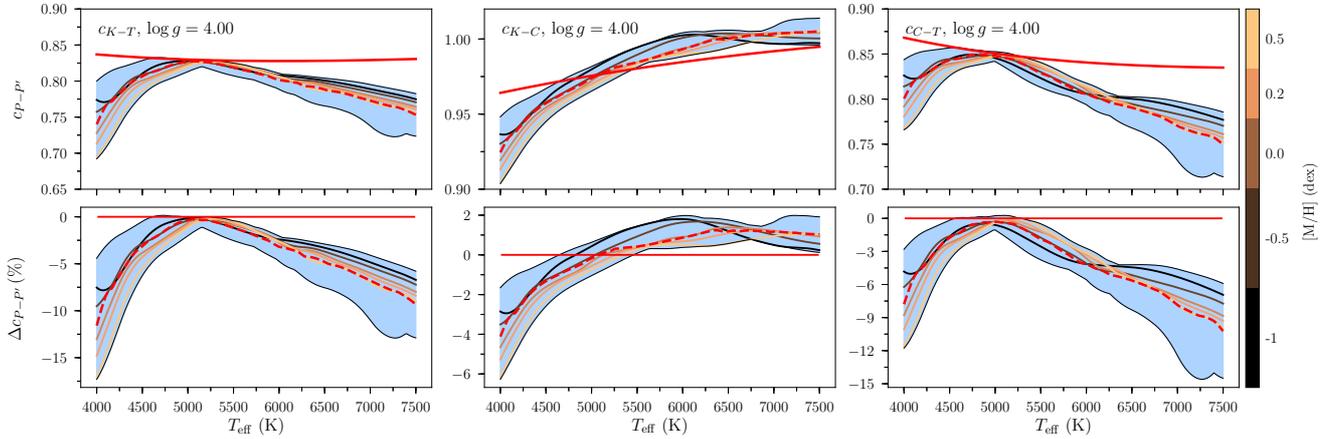}
\caption{Ratio between the bolometric corrections from \fref{fig:Bol_from_spec_ex} for \kp and TESS ($c_{K-T}$; left column), \kp and \co ($c_{K-C}$; middle column), and \co and TESS ($c_{C-T}$; right column). The lines coloured according to the color-bar give $c_{P - P'}$ for a specific \logg and a range of \meh values. In all panels the full red line gives the $c_{P - P'}$ value from assuming a Planck spectrum (see \fref{fig:relation}); the blue shaded area shows the regions covered by all combinations of \logg and \meh, and the dashed red line gives the median $c_{P - P'}$ value for all these combination. The bottom row panels show the percentage difference between the Planck and synthetic spectrum $c_{P - P'}$ values.}
\label{fig:Bol_from_comp}
\end{figure*}
There is furthermore a dependence on the $\alpha$-enhancement of the star (see \sref{sec:alpha}), following the same tendency as \meh, \ie, that an increased $\alpha$-enhancement decreases the \cp-value. For most \teff this would increase the difference to the Planck-derived values.

In the online material of this paper we provide the full table of calculated \cp-values. The table is accompanied by a small piece of \texttt{Python} code that allows for an easy interpolation in the \cp grid.

We also tried calculating \cp using the high resolution synthetic spectra from the library of \citet[][]{2013A&A...553A...6H}\footnote{\url{http://phoenix.astro.physik.uni-goettingen.de/}} using the PHOENIX code \citep{1995ApJ...445..433A}. However, the results from these spectra showed an erratic behaviour with fluctuations in \cp of up to $\pm50\%$ from the Planck values in \teff spans of ${\sim}100$ K, and with the position in \teff and amplitude of these fluctuations varying with \logg and \feh. We could trace the variability back to strong fluctuations in the \teff-derivative of the spectral flux at a given wavelength. Until this behaviour is better understood we provide only results from ATLAS9, and caution that the choice of model atmosphere used for calculating \cp can have an impact on the results.


\subsection{Effects of interstellar extinction}
\begin{figure*}
    \centering
    \subfloat{\includegraphics[width=0.48\textwidth]{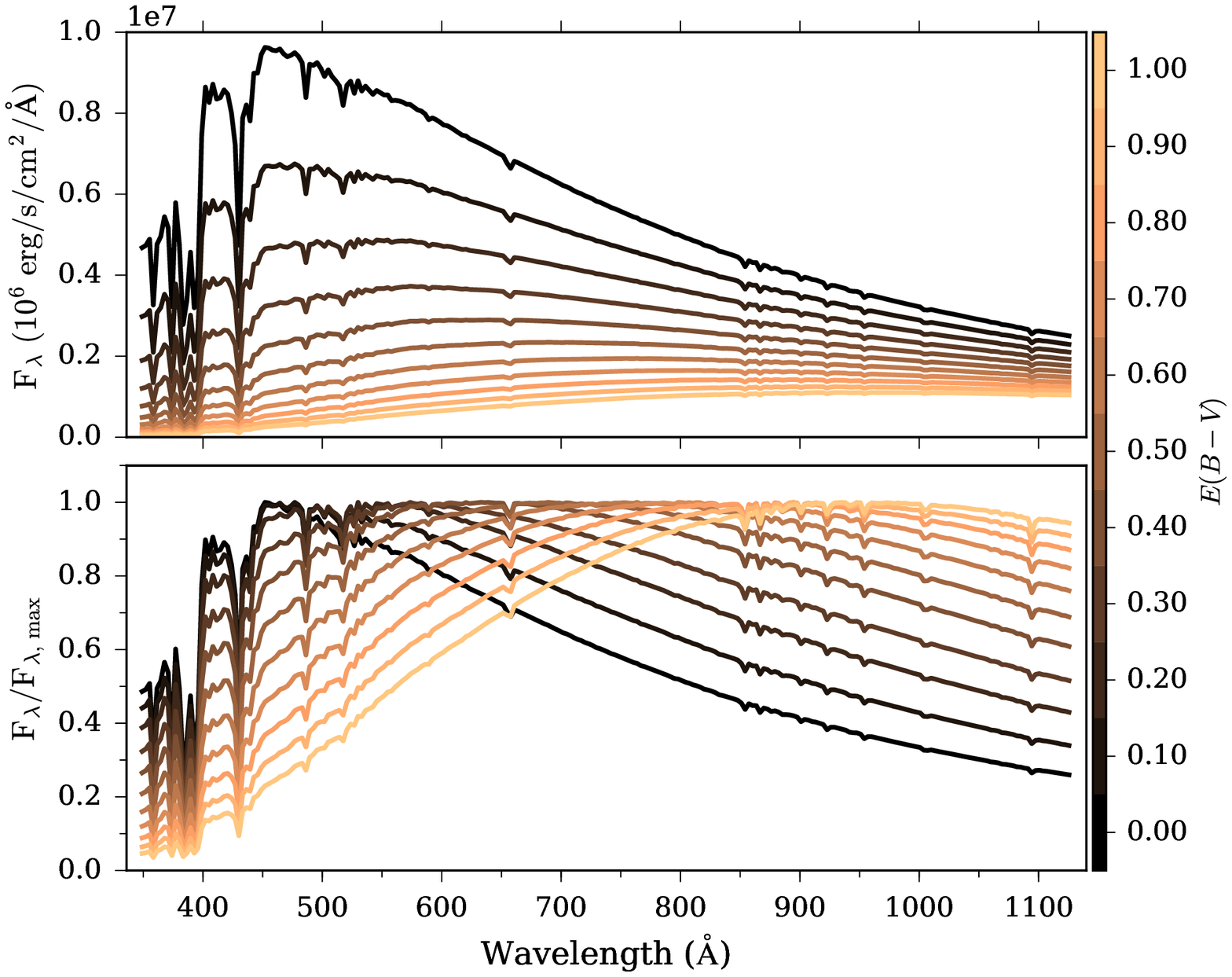}}
    \hfill
    \subfloat{\includegraphics[width=0.48\textwidth]{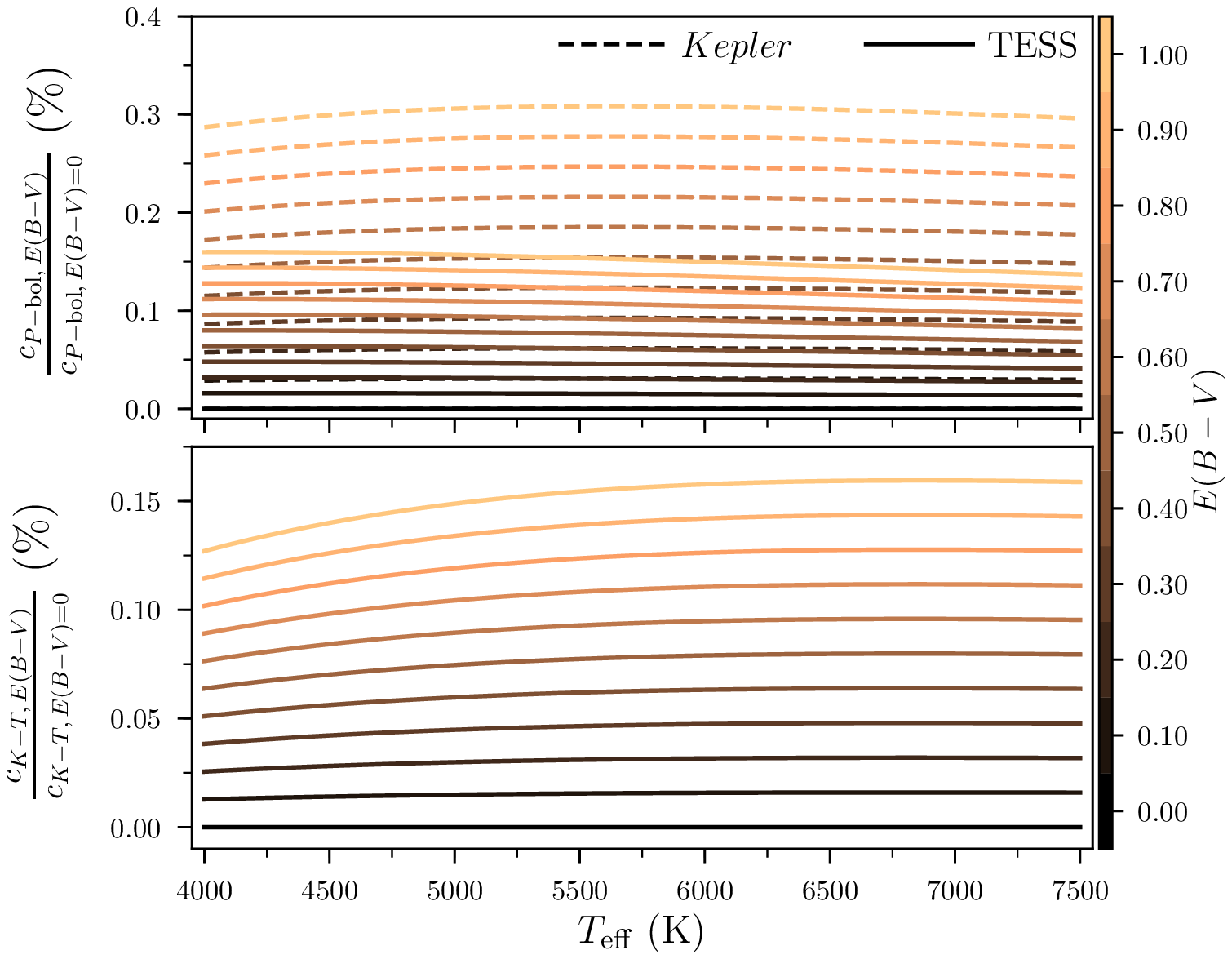}}
\caption{Left: Impact (\eqref{eq:ext}) from different levels of extinction (see colour bar) on the synthetic spectrum of a star with $\teff=5700$ K, $\logg=4.5$ dex, and $\meh=0$ dex. The bottom panel shows the top panel spectra normalised to their maximum flux to show at which wavelength the impact is the largest in relative terms. Right: relative change in \cp-values as a function of \teff from different values of $E(B-V)$ for the \kp and TESS missions (\co is similar to \kp). The bottom panel shows the corresponding relative change in the ratio $c_{K-T}$.  }
\label{fig:red}
\end{figure*}
Interstellar extinction from dust and gas along the line-of-sight will reduce the amount of stellar flux received, hence change the shape of the stellar spectrum. The extinction affect the spectrum differently at different wavelengths, with shorter wavelengths being more affected than longer ones \citep[][]{1930PASP...42..267T}, thereby causing a reddening (see \citet[][]{2003ARA&A..41..241D} for a review). To test the impact on \cp from reddening we use the $A_{\lambda}/E(B-V)$ relation from \citet[][]{1999PASP..111...63F} for the Galactic average value of $R_V=3.1\, [\equiv E(B-V)/A_V]$ \citep[][]{1989ApJ...345..245C,1992A&AS...93..449B,2007ApJ...663..320F}.
The range of wavelengths covered by either the \kp, \co, or TESS spectral responses cover a range of wavelengths where $A_{\lambda}$ changes fairly smoothly, and agree well with most determinations from other authors.
We note that the value of $R_V$ will depend on the specific line-of-sight to the star, but for the wavelength range covered here the dependence of $A_{\lambda}/E(B-V)$ on $R_V$ is quite modest and should not be a main source of uncertainty.

For a given $E(B-V)$ we apply the extinction to the flux, $F_{0,\lambda}$, from the stellar spectra as
\begin{equation}\label{eq:ext}
F_{\lambda} = F_{0,\lambda}\times 10^{-0.4\, A_{\lambda}}\, .
\end{equation}

The left panel of \fref{fig:red} shows the effect of applying the wavelength-dependent extinction from a range of $E(B-V)$ reddening values to the spectrum of a $\teff=5700$ K, $\logg=4.5$ dex, and $\meh=0$ dex star. The overall reddening (\ie, move to longer wavelengths) is seen, and also indicate that stars of different \teff should be affected to different extents, with a larger relative change the bluer (hotter) the star. Any effect should therefore also be larger for the \kp/\co band-pass as compared to that of TESS, because the former is slightly bluer (see \fref{fig:Transfer_functions}). 

The right panel of \fref{fig:red} shows the relative change in \cp from adopting different values of $E(B-V)$. While the impact from reddening follows expectations in terms of \teff and band-pass, it is seen that the impact is very modest and at the sub-percent level. This can be understood by the fact that the influence on the spectrum also cause a change in the spectrum derivative, resulting in only a small net effect on the integral of their ratio in \eqref{eq:cbol}. So, in addition to having a negligible effect on measured oscillation amplitudes in a given band-pass, interstellar extinction is also of little importance in the conversion to bolometric amplitudes and can largely be ignored.

\subsection{Effects of stellar parameter uncertainties}\label{sec:unc}
\begin{figure*}
    \centering
    \subfloat{\includegraphics[width=0.47\textwidth]{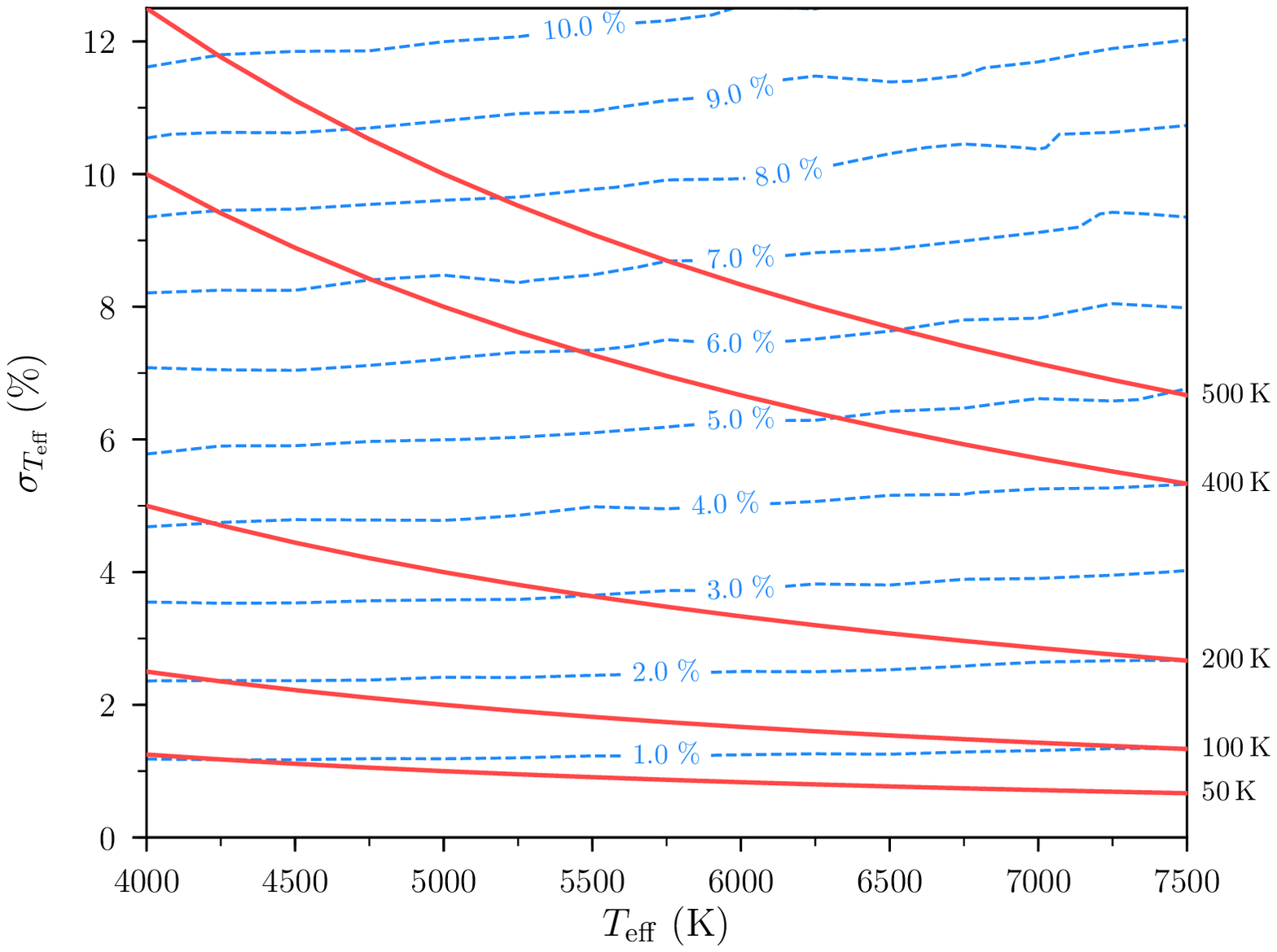}}
    \hfill
    \subfloat{\includegraphics[width=0.47\textwidth]{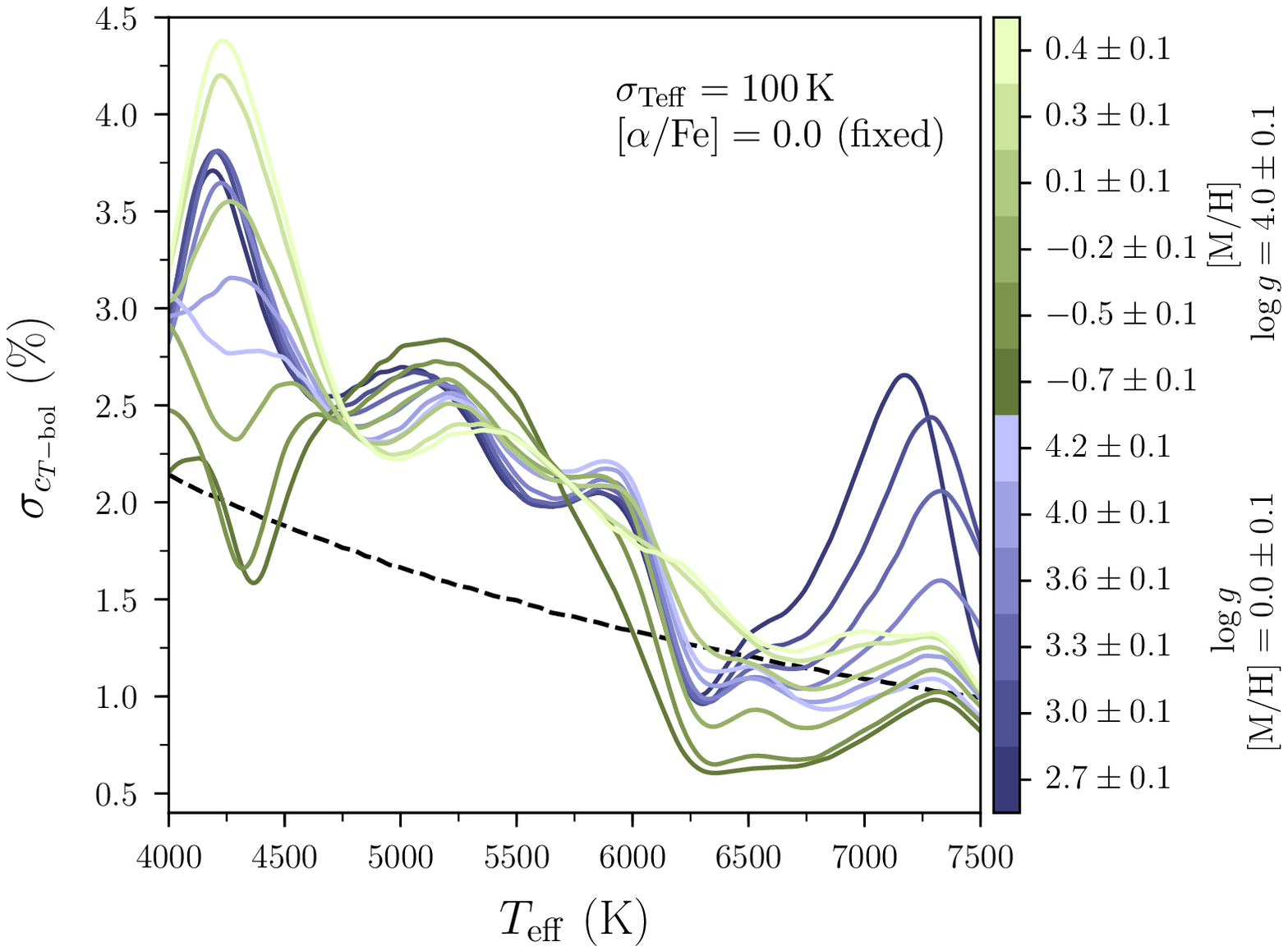}}
\caption{Left: Relative uncertainty on $c_{T\rm -bol}$ (TESS) as calculated from a Planck spectrum represented by blue dashed contour as a function of \teff and the relative uncertainty on \teff ($\sigma_{\teff}$). Indicated are also lines of constant absolute \teff uncertainty, with values indicated on the right side of the figure. Right: Relative uncertainty on $c_{T\rm -bol}$ (TESS) as calculated from synthetic spectra as a function of \teff. For comparison we also indicate the line of $c_{T\rm -bol}$ uncertainty based on Planck spectra (see left panel), where the uncertainty only depend on \teff. The colour-bar is split in two, with blue colours indicating varying \logg with a fixed \meh, and vice versa for the green colours.}
\label{fig:uncertainty}
\end{figure*}

When using a bolometric correction on observed amplitudes it is naturally important to propagate any uncertainty in parameters upon which \cp depends. For the corrections based on Planck spectra only the \teff uncertainty needs to be taken into account. \fref{fig:uncertainty} (left panel) give contours for the relative uncertainty on $c_{T\rm -bol}$ as a function of \teff and its relative uncertainty. As seen, the contours for the relative uncertainty on $c_{T\rm -bol}$ are nearly horizontal in the plot, \ie, it is nearly constant for a given relative uncertainty on \teff. It is however slightly smaller -- this is expected given the reasonable representation of the $c_{T\rm -bol}$ dependence on \teff by a simple power-law (\eqref{eq:rel1}), in which case the relative uncertainties should scale simply with the exponent ($\alpha$) of the power-law.
In terms of which parameterization to use for the Planck calculated values (Eqs.~\ref{eq:rel1}-\ref{eq:rel2}; \tref{tab:rel_coeff}), it is useful to take the effect of the temperature uncertainty into consideration. Taking the example of a star with temperature $\teff=5777\pm 80$ K, the corresponding bolometric correction for TESS would be $c_{T\rm -bol}=1.18\pm 0.01$. The uncertainty on the temperature of $\pm80$ K thus gives an uncertainty on $c_{T\rm -bol}$ that is a factor of ${>}4$ times larger than the $\sigma_{\rm rms}$ for the simplest $T_1$ parameterization (\eqref{eq:rel1}). Hence, even for the simple $T_1$ model, any bias will typically be well below the uncertainty on $c_{T\rm -bol}$ from \teff alone.

To appraise the impact on the uncertainty of $c_{T\rm -bol}$ from having dependencies on \logg and \meh we performed a Monte Carlo sampling in the grid of calculated $c_{T\rm -bol}$ values, including normally distributed uncertainties on \teff, \logg, and \meh. The sampling was done by first tessellating the $c_{T\rm -bol}$ grid using a Delaunay triangulation, followed by a linear barycentric interpolation on the triangulation simplices. \fref{fig:uncertainty} (right panel) shows the derived relative uncertainty from this exercise when adopting an uncertainty on \teff of $100\, \rm K$. As expected, the dependence of the relative uncertainty on \teff shows more structure, but generally follow the same trend as the corresponding Planck-derived uncertainty. In any case, a calculation of the uncertainty on \cp should be derived following an approach as the one described above for any specific use. The code accompanying the online grid of \cp values allows for such a calculation.


\section{Discussion}\label{sec:dis}

\subsection{Scaling relations}\label{sec:scale}
To date many empirical relations exist to describe either amplitudes of solar-like oscillation modes \citep[][]{1995A&A...293...87K,2011ApJ...743..143H,2011ApJ...737L..10S,2013MNRAS.430.2313C} or the surface granulation \citep[see, \eg,][]{2011ApJ...741..119M, Kallinger2014}.

If such a scaling relation is intended to enable a precise representation of amplitudes for a given photometric mission, then the bolometric correction is of little importance and the dependencies of the amplitude (with \teff, \logg, \meh, etc.) caused by the specific band-pass can be incorporated in the representation \citep[see, \eg,][]{Yu2018}.

A proper description of \cp becomes important, however, if the scaling relation is meant to enable a comparison of measured amplitudes with theory and to uncover dependencies on fundamental stellar parameters of the driving of solar-like oscillations. The important point is, that if scaling relations are derived for bolometric values, \ie, based on values converted from the measured and band-pass specific values, then one should be careful in interpreting any dependency with stellar fundamental parameters as an actual properties of the underlying driving mechanism if the dependencies belonging to the conversion itself have not been accounted for. The same goes for relations describing the dependence of granulation amplitudes on stellar parameters.

Conversion to bolometric amplitudes for \kp data for generating scaling relations and testing dependencies on, \eg, metallicity, have so far made use of the \citet{B11} conversion \citep[see, \eg,][]{2011ApJ...743..143H,2012A&A...537A..30M,2013MNRAS.430.2313C,Kallinger2014,Vrard2018}, as have some comparisons with theory \citep[\eg][]{2013A&A...559A..40S}. In addition to generally resulting in underestimated uncertainties on the bolometric amplitudes, this conversion also neglect correction-specific dependencies on metallicity and surface-gravity. 

In recent years several authors have addressed the dependence of oscillation and granulation amplitudes on metallicity, a dependence that has been suggested by theoretical studies such as \citet{Houdek1999} and \citet{Samadi2010}. \citet{Vrard2018} investigated the metallicity dependence of bolometric oscillation amplitudes obtained from \kp and found a higher-than-20$\%$ change in amplitude from a 1 dex change in metallicity. This is in qualitative agreement with earlier findings from \citet{2012A&A...537A..30M}, who like \citet{Vrard2018} adopted the \citet{B11} conversion. \citet{Corsaro2017} looked at amplitudes of meso-granulation as measured in the \kp band-pass and found a significant dependence on metallicity. However, by working with \kp-specific amplitudes, rather than bolometric values, the measured dependence on metallicity will be a combination of the star and the band-pass dependencies, rather than being solely from the star. 
\begin{figure}
\includegraphics[width=\columnwidth]{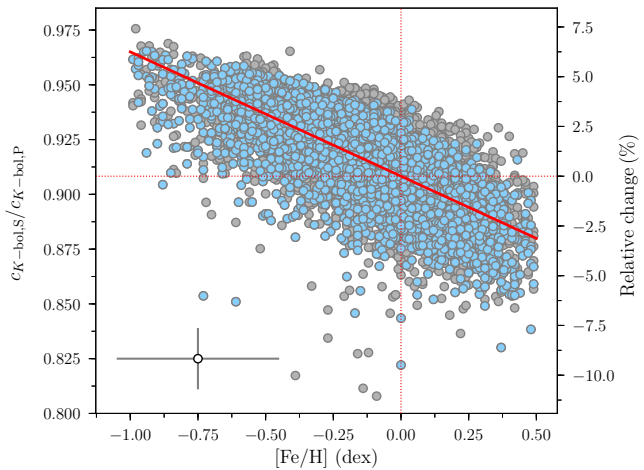}
\caption{Ratio between the bolometric correction \cp for the \kp mission as calculated adopting synthetic model spectra over the Planck spectrum, as a function of metallicity. \cp-values are computed using the Monte-Carlo sampling approach described in \sref{sec:unc}. The point in the lower left corner illustrate the median uncertainties of the sample. Ratios are presented for the sample of red giants with amplitudes measured by \citet{Yu2018}, but limited to stars identified as being in the Helium core burning phase. Points coloured blue are restricted further to a \logg range from 2.35 to 2.45. The red line gives the linear fit to the blue points from an orthogonal distance regression. The axis on the left side of the plot gives the change relative to an intersection at $\feh=0$.}
\label{fig:amprel}
\end{figure}
It is beyond the scope of this paper to assess the detailed impact on the reported dependencies of, especially, metallicity on oscillation and granulation amplitudes. However, to get a feel for the potential scale of the impact we considered the sample of amplitudes from \citet{Yu2018}, with values for \teff and \feh adopted from \citet{Mathur2017}. We can write bolometric amplitudes obtained from using \cp values from synthetic spectra in terms of those from using the Planck correction as (here for \kp):
\begin{equation}\label{eq:crat}
    A_{K\rm -bol, S} = {A}_{K\rm -bol, P}\,  \frac{c_{K\rm -bol, S}}{c_{K\rm -bol, P}}\, ,
\end{equation}
hence, any dependence of $A_{K\rm -bol, S}$ on \feh would be given by the dependence of $A_{K\rm -bol, P}$ on \feh \citep[\eg][]{Vrard2018}, but modified by the \feh dependence of the ratio of \cp values. We may therefore simply look at the \feh dependence of the \cp ratio.
For the calculation of \cp values we use the MC method described in \sref{sec:unc}. Rather than using the value for \logg provided by \citet{Yu2018} we re-calculate these (using the same method as \citet{Yu2018}) to include the inherent correlation between the \logg and \teff values from this method. To limit the range of \logg values we further restrict the sample to stars in their Helium burning phase with \logg from 2.35 to 2.45.
\fref{fig:amprel} shows the dependence of the calculated ratios with \feh. As seen from the fitted relation (red line) the ratios decrease by ${\sim}6.25\%$ from a 1 dex increase in \feh. From \eqref{eq:crat} we see that such a decrease in the \cp ratio with \feh would act against the increase in $A_{K\rm -bol, P}$ with \feh seen, \eg, by \citet{Vrard2018}. 

This exercise is by no means meant to suggest that dependencies of bolometric amplitudes reported in the literature are incorrect. We do, however, caution that part of the observed dependencies may be caused by applying a bolometric correction that neglects the dependencies under study. For comparisons with theoretical predictions we recommend either adopting a bolometric correction incorporating \feh and \logg dependencies, or even better, to apply the spectral response of the given mission of comparison to the theoretical calculations \citep[see, \eg,][]{2011ApJ...741..119M}.

\subsection{Detectability prediction}
\label{sec:det}

The importance of \cp enters if one requires a conversion from one band-pass to another. A direct implication from a change of the bolometric correction in this respect comes when predicting the detectability of solar-like oscillations. When preparing target-lists for observations in a photometric mission, such as TESS, the detectability of solar-like oscillations is assessed and targets reaching a certain likelihood of detection are proposed for observations. For TESS, targets that require observations with a 2-min cadence must be proposed, while targets that can be studied with the long-cadence option (currently a 30-min cadence) will be observed in any case from the TESS full-frame images. In the detectability calculation an estimate is needed for the amplitude of the oscillation modes \citep[see, \eg,][]{2011ApJ...732...54C}, which is typically obtained from scaling relations \citep[see, \eg,][]{1995A&A...293...87K,2011ApJ...743..143H,2011ApJ...737L..10S,2013MNRAS.430.2313C}.  

For solar-like stars to be observed with TESS, \citet{Campante2016} performed a detectability study following the procedures of \citet{2011ApJ...732...54C}. In their calculations an average amplitude conversion factor from \kp to TESS ($c_{K-T}$; \eqref{eq:conv}) of 0.85 was adopted, and this is also used for generating the actual target proposal list \citep{2019arXiv190110148S} -- while being slightly larger, this corresponds roughly to the level found here from the Planck-spectrum calculations (see \fref{fig:relation}).
To estimate the potential impact of changing the value of $c_{K-T}$, we can use the ratio of the signal-to-noise ratios ($S/N$; oscillation power over the stellar and instrumental noise background) from a specific $c_{K-T}$ value compared to the value of $0.85$:
\begin{equation}
R_{SN}\equiv \frac{S/N_{c_{K-T}=0.85}}{S/N_{c_{K-T}}} \approx \left(\frac{0.85}{c_{K-T}}\right)^2 \frac{b_{\rm inst} + (c_{K-T})^2 P_g}{b_{\rm inst} + (0.85)^2 P_g},\,
\end{equation}
where $b_{\rm inst}$ gives the instrumental noise, and $P_g$ gives the power level from the granulation background. Given that $c_{K-T}$ calculated from synthetic spectra in all cases are below $0.85$ (see \fref{fig:Bol_from_comp}), $R_{SN}$ will maximally reach a value of $(0.85/c_{K-T})^2$. If we adopt a value of $c_{K-T}\approx 0.8$, which corresponds to the value for a star with a temperature of ${\sim}6500\, \rm K$ and in general one of the lowest $c_{K-T}$ values for stars that will be observed in short-cadence ($2$-min) mode. Using a description of $b_{\rm inst}$ following \citet{2015ApJ...809...77S} and $P_g$ following \citet{Kallinger2014} (in the form used in \citet{Campante2016}) we find that for a star with a value of \numax of ${\sim}800\,\rm\mu Hz$, $R_{SN}$ reaches the maximum value of ${\sim}1.13$, which then decrease as a function of TESS magnitude.

The current target selection strategy is somewhat conservative in the sense of including, with low priority, some stars that are not expected to show oscillations. The up-to ${\sim}13\%$ overestimate of the predicted $S/N$ from using a fixed $c_{K-T}=0.85$ is unlikely to cause any significant change to the targets selected. 

\subsection{Other considerations}
\label{sec:other}

There are a number of effect that can impact the effective band-pass of the observations, hence the bolometric correction. While these effect are typically expected to be of little importance they are worth keeping in mind when processing photometric data for the sake of measuring amplitudes. We will not attempt a quantitative assessment of the potential impact of these effect. Below we will focus on effects particular to the TESS mission. 
\begin{itemize}
    \item The backside-illuminated CCDs of TESS are at the bottom equipped with high-conductivity ``straps'' for a fast frame-transfer. The conductive straps have a wavelength-dependent reflectively which takes effect in the IR (above ${\sim}800\,\rm nm$) where photons can pass through the entire silicon layer of the CCD. This effectively increase the QE, hence the SRF (\fref{fig:Transfer_functions}), of the pixels affected by the straps with an intensity rise of $0.5-14\%$ over wavelengths from $825-1050$ nm \citep[see][]{Kris2017}. For more information on the conducting (and other types of) straps we refer to the TESS Instrument Handbook \citep{TESSInstrumentHandbook}. 
    \item The QE of the TESS CCDs is mildly sensitive to temperature \citep[see][]{Kris2017}. While this is likely a negligible effect it could introduce a systematic change in the bolometric correction with a change in the CCD temperature, \eg, when the Earth is in/near the FOV.
    \item The Point Spread Function (PSF) of stars observed with TESS will include both off-axis and
chromatic aberrations arising both from the refractive elements of the TESS camera and from the deep-depletion CCDs, absorbing redder photons deeper in the silicon \citep{woods2016}. It has also been observed that for some saturated stars diffuse vertical and horizontal image extensions appear, known as ``mustaches''. These originate from reflections of long-wavelength light within the bulk silicon of the CCDs and will be made of photons with wavelengths ${>}950$ nm \citep{TESSInstrumentHandbook}. With these effects in mind it is important to use apertures that capture the full light from the star. 
\end{itemize}


\section{Conclusion}\label{sec:con}
We have calculated the bolometric correction factor \cp for the \kp, \co, and TESS missions, enabling the conversion of measured amplitudes for radial ($l=0$) oscillation modes and granulation to bolometric values. The numerical calculations followed the prescription set forth by \citet{B11} and \citet{M09} for the \kp and \co missions. 

The bolometric correction was computed using both a Planck spectrum and synthetic model flux spectra from ATLAS9 as representations for the stellar spectral flux density. For the \cp values adopting a Planck spectrum (\sref{sec:bol2}) we derived simple power-law (\eqref{eq:rel1}) and polynomial parameterizations (\eqref{eq:rel2}) as a function of \teff (\fref{fig:relation}), the coefficients of which are provided in \tref{tab:rel_coeff}. For the \cp calculations adopting synthetic spectra (\sref{sec:bol3}) we provide online tables of the values as a function of \teff, \logg, \feh, and $\alpha$-enhancement (\sref{sec:alpha}).

We find that adopting synthetic model spectra can change the bolometric correction by as much as $30\%$ at the lowest temperatures ($4000$ K), while the difference typically lie within ${\sim}5-10 \%$ around a \teff of $5500-6000$ K. After \teff, the metallicity has the strongest influence on \cp, followed by \logg. We find that amplitude comparisons between the missions included in this study can be off by up to $15\%$ when assuming a black-body spectrum over a more realistic flux spectrum.

The conversion of measured amplitudes to bolometric ones enabled by the \cp values can be used to explore dependencies of amplitudes on fundamental stellar parameters for a better understanding of the turbulent convection driving the excitation. We find that part of the dependence of stellar oscillations amplitudes on \feh reported in the literature may be caused by adopting \cp values that neglect this dependence (\sref{sec:scale}).

The \cp values can similarly be used for conversion of amplitudes observed by different photometric missions --- this can be used to assess the detectability yield of solar-like oscillations for a mission like TESS given amplitude relations from \kp, hence help guide the target selection. For the TESS mission, we find that a change of the adopted \cp values from the current ones assuming a stellar black-body spectrum would only make the target selection strategy more conservative (\sref{sec:det}).

The use of synthetic spectra from 1D model atmospheres provide an improvement over the Planck spectrum in terms of appreciating dependencies on parameters other than \teff. They are, however, still limited by neglecting potential influences from spectral line formation in non-local thermodynamic equilibrium, and effects from a non-stationary 3D time-dependent modelling of the convective flows in the stellar atmosphere. A future advancement would be to compute \cp values from the grid 3D model atmospheres from, for instance, the Stagger project \citep{stagger1,stagger2}. 

In this study we tried also computing \cp using a spectral library adopting the PHOENIX code. The results from this were very erratic and will not be provided until the cause is better understood. We caution that the choice of model atmosphere will likely have some influence on the derived results, as was found for oscillation mode visibilities by \citet{B11}. 

We will in a future work apply the same synthetic spectra used for the bolometric correction to the calculation of oscillation mode visibilities \citep[see][for such calculations for \kp]{B11}. Once the TESS mission has observed the \kp field during its second year of operation, we will test the values provided in this work by comparing amplitudes measured by both missions. The empirical determination of the \cp-ratios between \kp and TESS will also be important to test different 1D atmosphere models, different implementations of model physics, and potential improvements from 3-D model atmospheres.


\section*{Acknowledgments}
\footnotesize
We thank the referee J\'{e}r\^{o}me Ballot for comments that helped improve the paper.
We thank William~J.~Chaplin and Guy~R.~Davies for useful comments on the paper, and Roland Vanderspek for discussions on the spectral response function of TESS. We thank Andrea Miglio and Ennio Poretti for supplying the \co spectral response function. MNL acknowledges the support of The Danish Council for Independent Research | Natural Science (Grant DFF-4181-00415), and support from the ESA PRODEX programme. MNL is grateful to the School of Physics and Astronomy at the University of Birmingham for their hospitality during parts of the study. Funding for the Stellar Astrophysics Centre is provided by The Danish National Research Foundation (Grant agreement no.: DNRF106).

\bibliographystyle{mnras}
\bibliography{TESS_bolometric_correction}

\appendix
\section{$\alpha$-enhancement}\label{sec:alpha}
\begin{figure*}
\includegraphics[width=\textwidth]{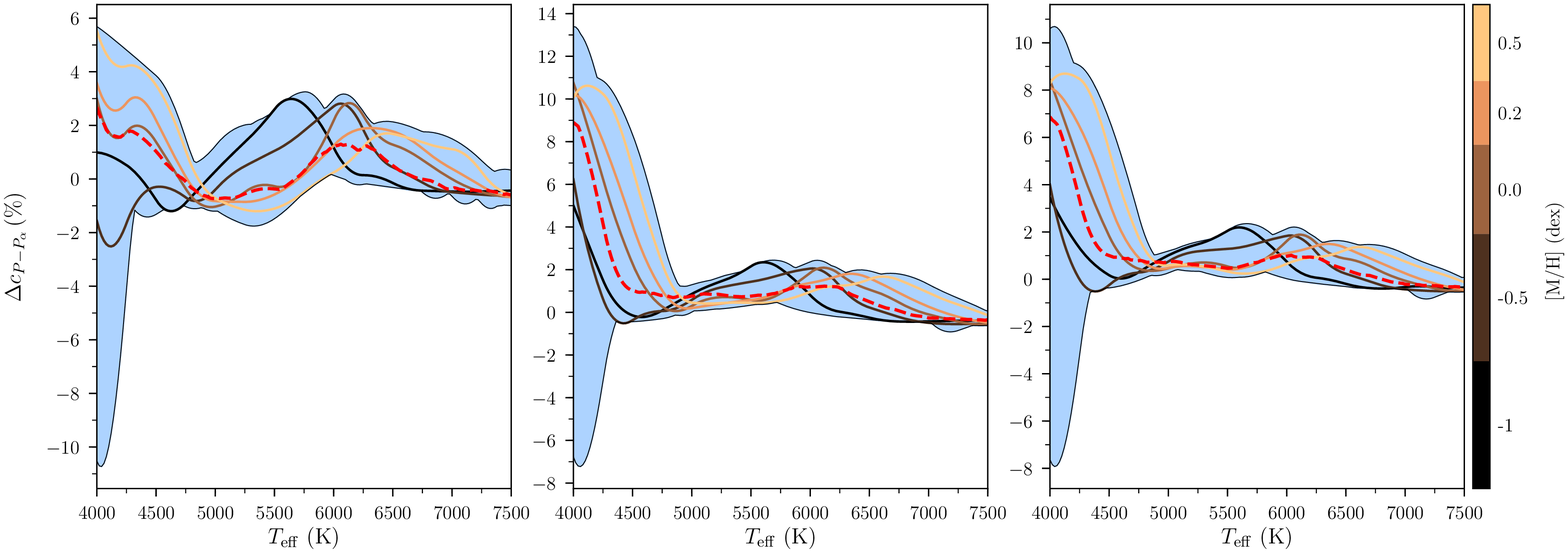}
\caption{Difference between the bolometric correction with $\alpha$-enhancement of $0$ and $+0.4$ for \kp (left), \co (middle), and TESS (right). The lines given are for $\logg=4.0$ and are colour-coded by metallicity; the shaded blue area shows the regions covered by all tested \logg-values. The dashed red line shows the median difference for all tested models, which in all cases is positive, \ie, \cp is lower for a higher $\alpha$-enhanced. }
\label{fig:Bol_from_spec_alpha}
\end{figure*}

The dependence of \cp on $\alpha$-enhancement is exemplified in \fref{fig:Bol_from_spec_alpha}, which shows the percentage difference from adopting an $\alpha$-enhancement of $+0.4$ as compared to a value of $0$ (used for \fref{fig:Bol_from_spec_ex}). As seen the effect is generally largest towards low \teff and high \meh, and overall the \cp values for the $\alpha$-enhanced case are lower than the non-enhanced -- this would for most \teff values increase the disagreement to the Planck-spectrum value.

\section{Relation parameters}
\tref{tab:rel_coeff} provides coefficients for the power-law (\eqref{eq:rel1}; $\alpha, T_o$) and polynomial relations (\eqref{eq:rel2}; $a_n, T_o$) describing \cp from assuming a Planck function as a function of \teff.
\begin{table*}
\centering 
\begin{threeparttable}
\caption{Parameters from the fits of Equations~\ref{eq:rel1} ($\alpha, T_o$) and \ref{eq:rel2} ($a_n, T_o$) to the numerically calculated \cp values from \eqref{eq:cbol} against \teff (\fref{fig:relation}). Corrections are given for \kp, \co, and TESS. The quality of the fits are quantified by the root-mean-square-error $\sigma_{\rm rms}$.} 
\label{tab:rel_coeff}
\begin{tabular}{ccccccc} 
\cmidrule[1.0pt](lr){1-7}\\ [-1.8ex]
Model &$T_o$ & $\alpha$ & $a_0$ &  $a_1$ & $a_2$ & $\sigma_{\rm rms}$\\ 
name & $\rm (K)$ &  & $\rm (K)$ &  $\rm (K^{-1})$ & $\rm (K^{-2})$ & \\ 
\cmidrule(lr){1-7}\\ [-1ex]
\multicolumn{7}{c}{$c_{K\rm -bol}(\teff)$ from \citetalias{B11}}  \\ 
\cmidrule(lr){1-7}\\ [-1.8ex]
$K_1$ & $5934$ & $0.80$		&  $\cdots$ 	& $\cdots$ 				& $\cdots$  				& $1.05\times 10^{-3}$    \\
$K_2$ & $5934$ & $\cdots$	&  1 		& $1.349\times 10^{-4}$ 	& $-3.120\times 10^{-9}$	& $7.24\times 10^{-5}$    \\
\cmidrule(lr){1-7}\\ [-1ex]
\multicolumn{7}{c}{$c_{C\rm -bol}(\teff)$ }  \\ 
\cmidrule(lr){1-7}\\ [-1.8ex]
$C_1$ & $5816$ & $0.74$		&  $\cdots$ 				& $\cdots$ 				& $\cdots$  	&$1.36\times 10^{-3}$	\\
$C_2$ & $5816$ & $\cdots$	&  $1.001$	& $1.295\times 10^{-4}$ 	&  $-3.288\times 10^{-9}$	& $3.81\times 10^{-5}$	\\
\cmidrule(lr){1-7}\\ [-1ex]
\multicolumn{7}{c}{$c_{T\rm -bol}(\teff)$ }  \\ 
\cmidrule(lr){1-7}\\ [-1.8ex]
$T_1$ & $4714$ & $0.81$		&  $\cdots$ 				& $\cdots$ 				& $\cdots$  	&$3.07\times 10^{-3}$	\\
$T_2$ & $4714$ & $\cdots$	&  $9.995\times 10^{-1}$	& $1.784\times 10^{-4}$ 	&  $-6.127\times 10^{-9}$	&$1.05\times 10^{-4}$	\\
\cmidrule[1.0pt](lr){1-7}\\ [-1ex]
\end{tabular} 
\end{threeparttable} 
\end{table*} 

\label{lastpage}

\end{document}